\documentclass[journal]{new-aiaa}
\usepackage[utf8]{inputenc}
\usepackage{float}
\usepackage{graphicx}
\usepackage{amsmath}
\usepackage[version=4]{mhchem}
\usepackage{siunitx}
\usepackage{longtable,tabularx}
\usepackage{subfig}
\title{Modeling and Bifurcation Analysis of Longitudinal Dynamics of an Air-Breathing Hypersonic Vehicle}

\author{Kavita Shekhawat\footnote{PhD Student, Department of Aerospace Engineering, IIT Madras, Student Member, AIAA.} and Nandan Kumar Sinha\footnote{Professor, Department of Aerospace Engineering, IIT Madras.}}
\affil{Department of Aerospace Engineering, IIT Madras, Chennai, 600036}

\begin{document}

\maketitle

\begin{abstract}
A nonlinear model of an Air-Breathing Hypersonic Vehicle (ABHV) longitudinal dynamics characterized by coupling of aerodynamic and propulsive terms is presented in this paper. The model is verified using modal analysis carried out around a design operating condition with results available in the literature. Further, parametric dynamic behavior is computed for the model as steady states with local stability with respect to its control inputs, elevator and fuel-equivalence ratio in four different cases using a numerical continuation algorithm. Detailed analysis of the qualitative longitudinal dynamics of the model is carried out based on bifurcation theory methodology. Numerical simulation results are presented to verify bifurcation analysis results.
\end{abstract}

\section{Nomenclature}

{\renewcommand\arraystretch{1.0}
\noindent\begin{longtable*}{@{}l @{\quad=\quad} l@{}}
$M_a$  & Mach number \\
$\alpha$ &    angle of attack (deg) \\
$\theta$& pitch angle (deg) \\
$Q$ & pitch rate (deg/s)\\
$\phi$ & fuel-equivalence ratio \\
$\delta_e$ & elevator deflection (deg) \\
$v$ &speed of sound \\
$L$ &lift \\
$D$   & drag \\
$M$  & pitching moment \\
$T$  & scramjet thrust \\
$\rho$  & air density \\
$m$  & vehicle mass 

\end{longtable*}}

\section{Introduction}
Hypersonic transport vehicle technology, with a history of over half a century, continues to remain a niche area of research interest with some successful demonstrations, such as NASA's X-43A. These vehicles powered by scramjet engines, broadly classified as Air-Breathing Hypersonic Vehicle (ABHV), offer distinct advantages compared to other types of aircraft. These vehicles can be cost-effective alternatives for accessing space and may also be useful for single-stage-to-orbit operations. Given their exceptionally high speeds and thrust dependent on angle of flow intake, it is crucial to understand the dynamics of these vehicles with respect to relevant parameters for quality and safety of operations\cite{1993schmidt_problems,1992schmidt_dynamics,1994chavez_analytical,2005bolender_non,2008torrez_effects,2014zhang_modeling,2015xu_overview}.

\begin{figure}[H]
\centering
\includegraphics[width=1.0\textwidth]{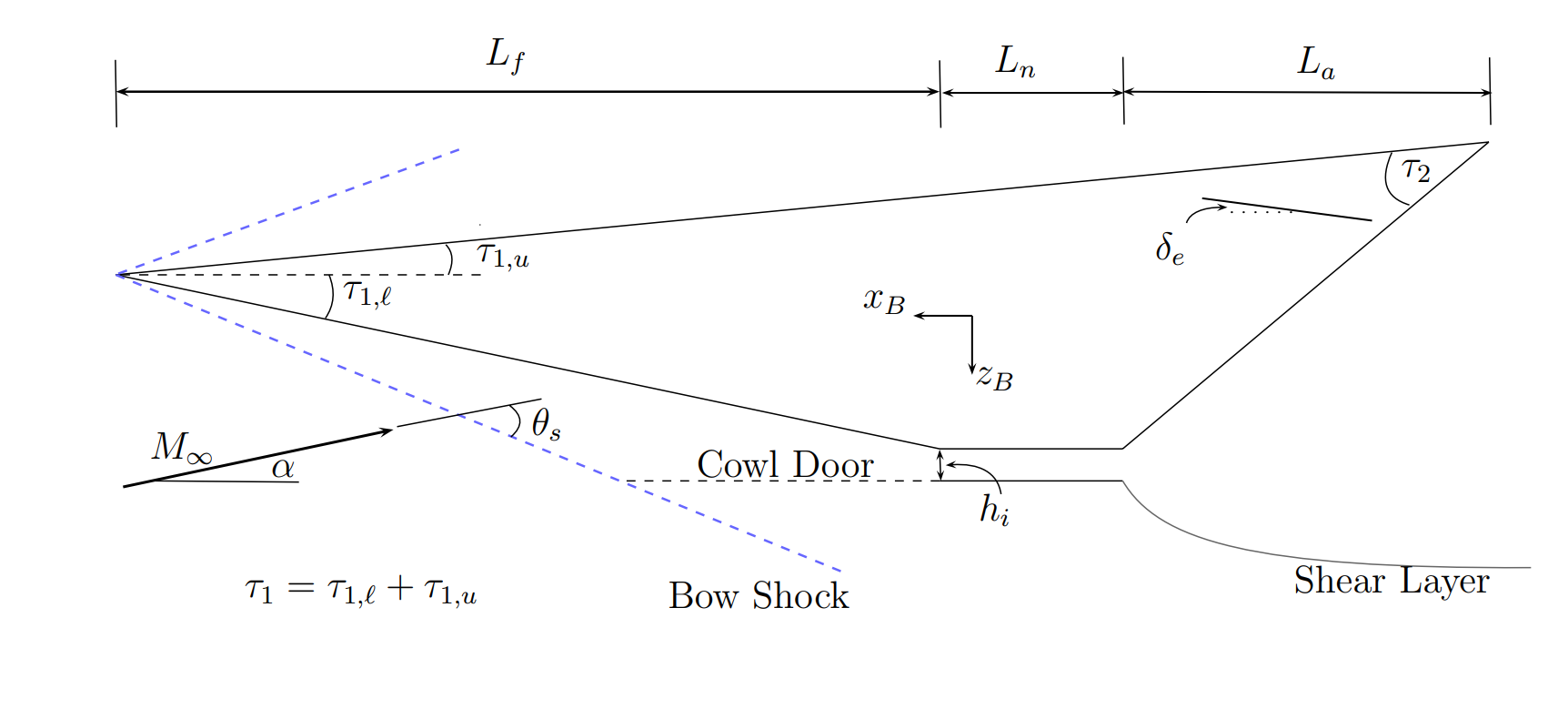}
\caption{Hypersonic Air-Breathing Vehicle Geometry\cite{2007bolender_nonlinear}.}
\label{fig:ABHV}        
\end{figure}

\begin{figure}
    \centering
    \includegraphics[width=0.55\textwidth]{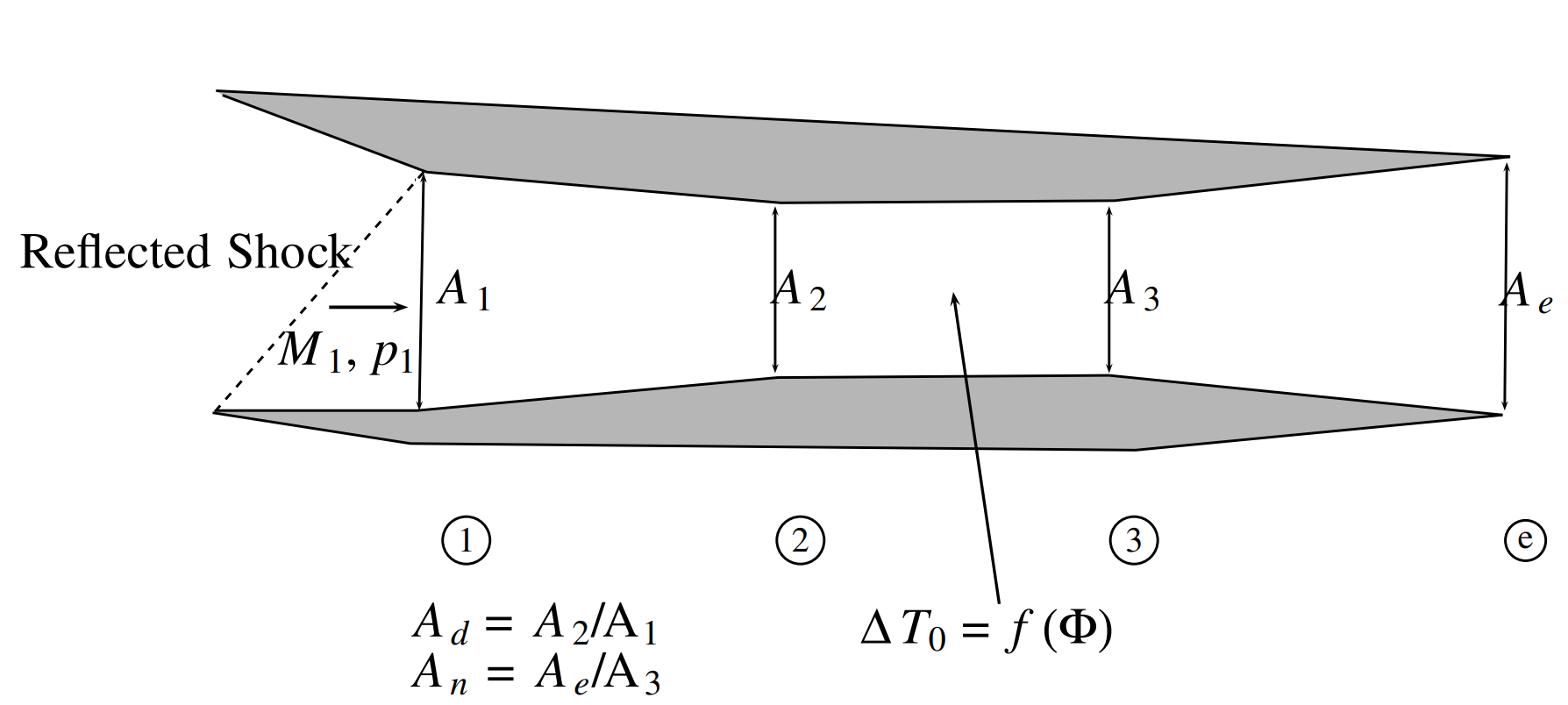}
    \caption{Scramjet Cross-section\cite{2007bolender_nonlinear}.}
    \label{fig:scram}
\end{figure}

An ABHV, as shown in Fig.~\ref{fig:ABHV}\cite{2007bolender_nonlinear}, is a long, slender lifting body with a sharp leading edge for drag reduction and an integrated underslung propulsion system with the airframe. The propulsion systems used for these vehicles relies on air-breathing scramjet engines as shown in Fig.~\ref{fig:scram}\cite{2007bolender_nonlinear},  which are specifically designed for efficient operation under hypersonic conditions. The angle-of-attack significantly influences airflow entering the scramjet engine, rendering thrust heavily dependent on it, sometimes also responsible for the unstart problem in flight. The underslung configuration of the scramjet engine induces a nose-up pitching moment directly proportional to thrust, creating a coupled interconnection between propulsion and aerodynamic effects\cite{2005bolender_non,2007bolender_nonlinear}, essentially making the model of an ABHV nonlinear. Flexibility effects due to the slender configuration of ABHV leading to structural deformation further add to the nonlinearities.
Over the past two decades, ABHV modeling has undergone significant advancements. Most of the vehicle dynamics models developed are primarily focused on nonlinear formulations that describe the behavior of the aircraft in longitudinal dynamics. In the literature, there have been several articles that discuss the challenges associated with the dynamics and control of ABHV. These include work by Chavez and Schmidt\cite{1992chavez_integrated,1992schmidt_dynamics,1993schmidt_problems,1994chavez_analytical}, with a first attempt at a comprehensive analytical model of the longitudinal dynamics of an ABHV. Another notable model proposed by\cite{1995bilimoria_integrated} encompassed additional factors such as rigid-flexible effects, the rotation of the Earth, wind, and various other influences. Bolender and Doman\cite{2005bolender_non,2007bolender_nonlinear}, expanded on previous work by incorporating compressibility accounting for viscous effects in their aerodynamic model. This advancement resulted in a more accurate and complex model compared to earlier efforts. This first-principle model (FPM), encompassing the longitudinal dynamics of a hypersonic air-breathing vehicle along with flexible dynamics, has become widely adopted in the literature of ABHVs. A more elaborate model including heave motion presented in\cite{2007parker_control} discussed additional couplings and increased fidelity.  

Subsequently, the focus shifted towards control-oriented modeling, which aims to facilitate the implementation of efficient control laws.  The first-principle model of the hypersonic air-breathing vehicle posed challenges for model-based control design due to its complexity. To address this, reduced-order models were derived from first-principle model to capture the vehicle’s dynamics while simplifying control design implementation\cite{2007parker_control,2011huifeng_control,2011falkiewicz_reduced,2018sudalagunta_aeroelastic}. One widely accepted control-oriented model (COM) is the one developed by Parker et al. \cite{2007parker_control}, known for its fidelity to the actual vehicle. The model also incorporated an actuator (canard), an additional design feature of the ABHV, to enhance the control authority of the vehicle required to compensate for the flexible effects.

Literature on Air-Breathing Hypersonic Vehicles (ABHVs) reveal that these aircraft are highly unstable and exhibit complex interactions between their subsystems\cite{1993schmidt_problems,2005bolender_non,2007parker_control}, prompting need of a thorough investigation into the nonlinear dynamics of ABHV model for safety and enhanced performance. To study phenomena arising out of nonlinearities, advanced tools such as bifurcation and continuation methods are not only necessary and efficient but also complement traditional linear model-based approaches \cite{1982carroll_bifurcation,2015sharmanumerical}. Singh et al.\cite{2022singh_longitudinal} used a bifurcation theoretic approach to study the dynamics of an ABHV model with thrust dependence on Mach number and altitude. Bifurcation analysis-based tools employing numerical continuation algorithm are highly efficient for investigating dynamical system models governed by parametrized ordinary differential equations. Advanced continuation algorithms, such as AUTO-07p\cite{2009Deodel}, generate a map of steady-state solutions as functions of parameters of nonlinear model of a vehicle including stability and bifurcations, thus revealing the global parametric dynamics of the vehicle. This technique is highly valuable for identifying regions where the dynamics are unacceptable and should be avoided, as well as for determining the magnitudes of disturbances that can be tolerated while maintaining stability. Over time, this method has evolved into a reliable tool for analyzing the global dynamics of nonlinear aircraft models. It has significantly advanced our understanding of various nonlinear dynamic phenomena in aircraft flight dynamics, including high angle-of-attack spin, deep stall, stall dynamics, wing rock, and inertia-coupled rolling motions\cite{sinha2021elementary,1980mehra_bifurcation,2012crespo_analysis,2020wu_analysis}. 
Further, bifurcation analysis approach has been used to
analyze aircraft dynamics in constrained flight conditions, such as, level flight, level turn flight, coordinated roll, etc., where multiple controls are deployed simultaneously\cite{2020wu_analysis,2001ananthkrishnan_level}. Computing envelope of achievable equilibrium states\cite{2006richardson_design}, understanding the nonlinear dynamic phenomena due to flexibility effects\cite{2001charles_bifurcation} and design of control laws\cite{2005paranjape_airplane,1997avanzini_bifurcation,2000ortega_bifurcation,1998littleboy_using} have been some other useful applications of bifurcation analysis-based methods in aircraft flight dynamics.

The paper is organized as follows. A nonlinear model of an Air-Breathing Hypersonic Vehicle (ABHV) longitudinal dynamics characterized by coupling of aerodynamic and propulsive terms is presented first in Section III. The model is verified using modal analysis carried out around a design operating condition with results available in literature. Brief introduction to bifurcation analysis methodology is presented in Section IV. Bifurcation analysis results for the ABHV model for four different cases are presented in
Section V followed by discussions, and concluding remarks are presented in Section VI.

\section{Air-Breathing Hypersonic Vehicle (ABHV) Model }
In this study, we consider a generic air-breathing hypersonic vehicle (ABHV), which is scramjet powered with an operating altitude of around 25-40 km. The geometry of the vehicle is shown in Fig.~\ref{fig:ABHV}\cite{2007bolender_nonlinear}. A control surface (elevator) is present to control primarily the pitching motion. It is represented as a flat plate that pivots at its midpoint accounting for shift of aerodynamic center to half-chord location at high speeds and allowing for full surface deflection. This design aligns with the X-43 and NASP configurations, where wingtips were deflected for pitch control. 

\begin{figure}[H]
    \subfloat[]{
        \includegraphics[width=0.45\textwidth]{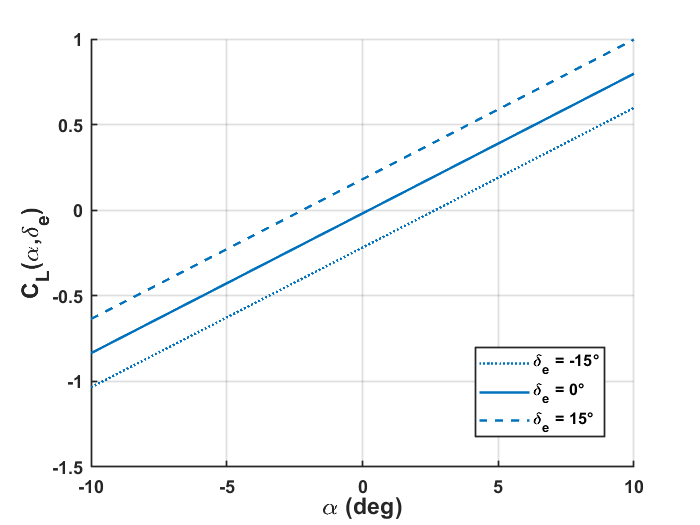}
        \label{fig:CL}
    }
    \hfill
    \subfloat[]{
        \includegraphics[width=0.45\textwidth]{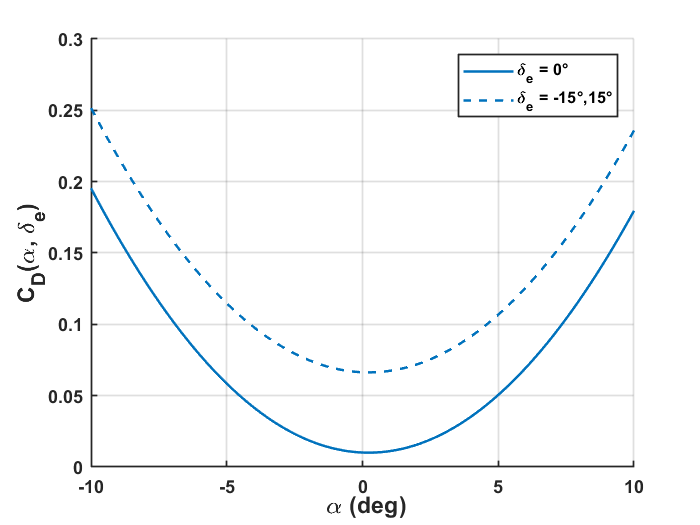}
        \label{fig:CD}
    }
    \hfill
    \subfloat[]{
        \includegraphics[width=0.45\textwidth]{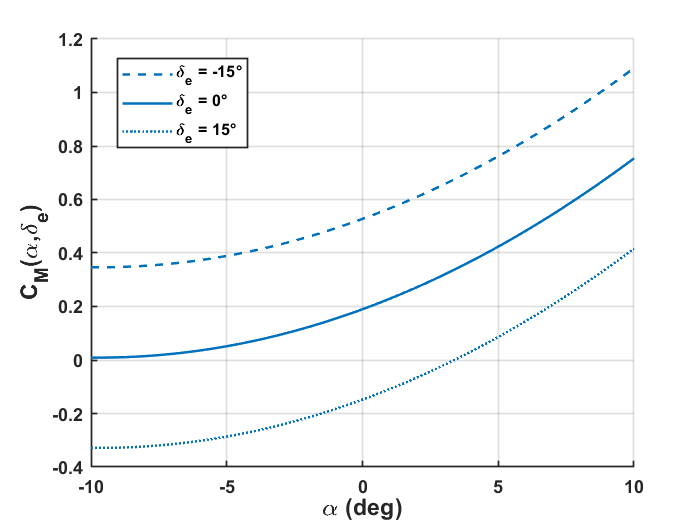}
        \label{fig:CM}
    }
    \hfill
    \subfloat[]{
        \includegraphics[width=0.45\textwidth]{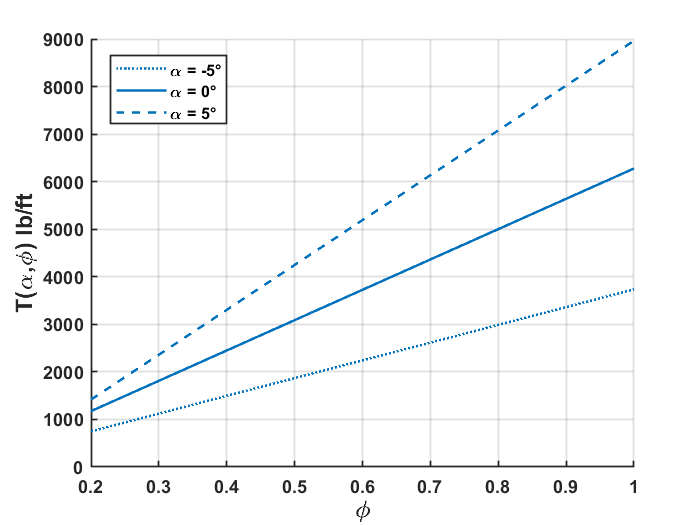}
        \label{fig:T}
    }
    \caption{Aerodynamic coefficients at three elevator settings and thrust variation for different values of angle of attack \textbf{($-5,0,5~\mathrm{deg}$)}.}
    \label{aerodata}
\end{figure}

\begin{figure}[H]
    \centering
    \includegraphics[width=0.6\textwidth]{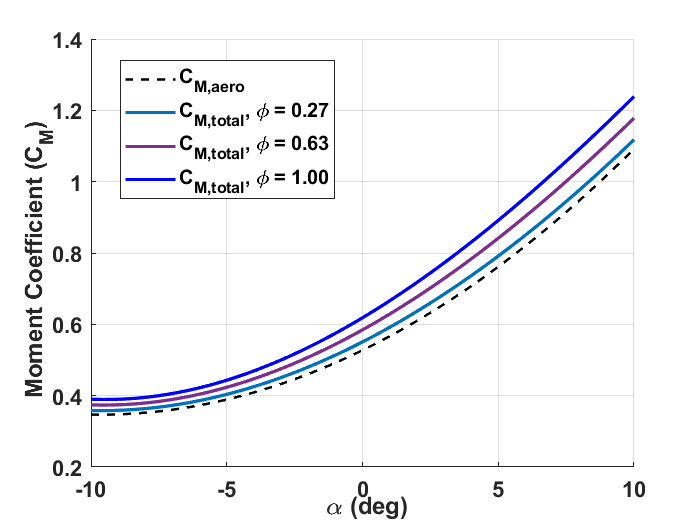}
    \caption{Total pitching moment coefficient consisting both aerodynamic and thrust components.}
    \label{fig:CMtotalvsaero}
\end{figure}

\begin{figure}[H]
\centering
    \subfloat[]{
        \includegraphics[width=0.47\textwidth, height=0.39\textwidth]{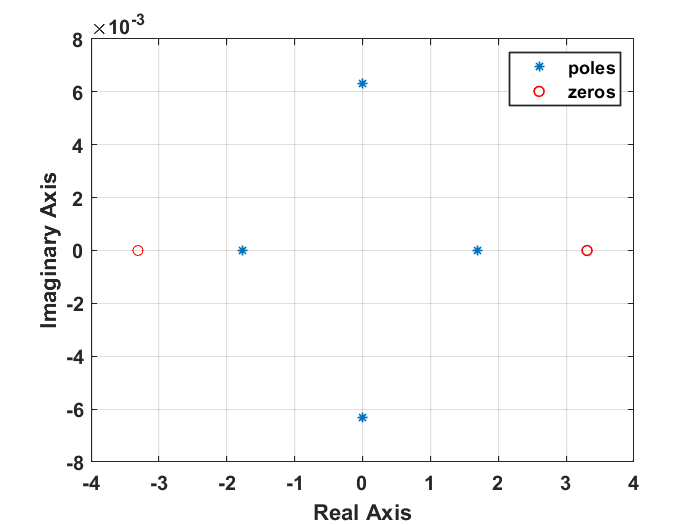}
        \label{fig:pole}
    }
    \hfill
    \subfloat[]{
        \includegraphics[width=0.5\textwidth]{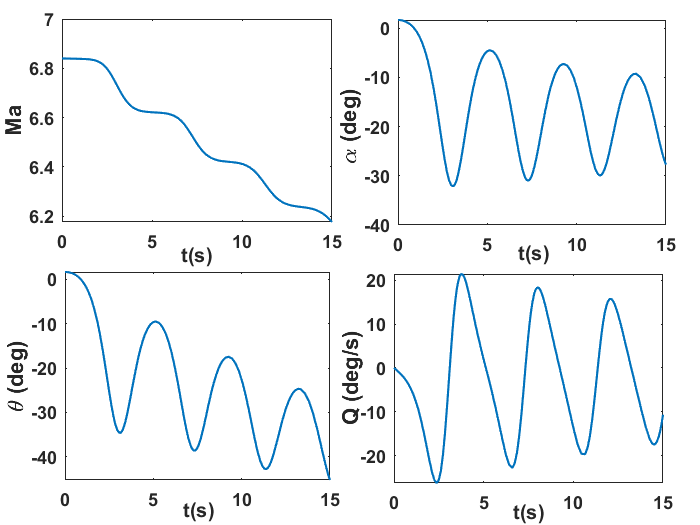 }
        \label{fig:OL15}
    }
    \caption{ (a) Pole/transmission zero maps of the Jacobian linearization of ABHV, (b) Open-loop time simulation around the nominal condition.}
    \label{polesimulation}    
\end{figure}

The other control input, the fuel-equivalence ratio ($\phi$), effectively controlling the fuel flow as shown in Fig.~\ref{fig:scram}\cite{2007bolender_nonlinear}, governs the change in total temperature resulting from the combustion process.
In general, thrust produced is a function of total temperature in the
combustor which is more strongly dependent upon fuel-to-air flow rates ratio ($f = \dot{m}_f/\dot{m}_a$) in the combustor. The fuel-equivalence ratio is defined as $\phi = f/f_{st}$, where $f_{st}$ is the stoichiometric fuel-to-air ratio of the scramjet fuel used (liquid hydrogen (LH2) or hydrocarbon based fuel such as JP-8 )\cite{2007bolender_nonlinear}. The rigid ABHV longitudinal dynamics model used in this study including the above control inputs is nonlinear due to state-dependent control inputs and complex interactions between the propulsion system and aerodynamics. Governing equations of motion of ABHV longitudinal dynamics are given as\cite{2007parker_control}:
\begin{align}
&\dot{M}_a =\frac{1}{v}\left[\frac{1}{m}(T \cos \alpha-D)-g \sin (\theta-\alpha)\right] \\
&\dot{\alpha}=\frac{1}{m(M_av)}\left(-T \sin \alpha-L\right)+Q+\frac{g}{M_av} \cos (\theta-\alpha) \\
&\dot{\theta} =Q \\
&\dot{Q}  = M/I_{y y} 
\end{align}
The system state vector $x_{T M} \in R^{4}$ is composed of the four state variables $M_a, \alpha, \theta, Q$. The control inputs $\delta_e$ and $\phi$ do not appear explicitly in the equations of motion. Instead, they enter through forces and moment terms.
The following expressions for force and moment are adopted from \cite{2007parker_control}:
\begin{align}
 & L=\frac{1}{2} \rho (M_av)^2 S C_L \\
 & D=\frac{1}{2} \rho (M_av)^2 S C_D \\
& M=z_T T+\frac{1}{2} \rho (M_av)^2 S \bar{c}\left[C_{M, \alpha}+C_{M, \delta e}\right] \\
& T=C_T^{\alpha^3} \alpha^3+C_T^{\alpha^2} \alpha^2+C_T^\alpha \alpha+C_T^0 
\end{align}
Here, the expression for $M$ includes an additional term $z_T  T$, where $z_T$ is a known value, to account for the pitching moment generated by the underslung scramjet engine. The analytical forms of the coefficients appearing in Eqs. (5 - 8) are \cite{2007parker_control}:
\begin{align}
& C_L=C_L^\alpha \alpha+C_L^{\delta e} \delta e+C_L^0 \\
& C_D=C_D^{\alpha^2} \alpha^2+C_D^\alpha \alpha+C_D^{\delta e^2} \delta e^2+C_D^{\delta e} \delta e+C_D^0 \\
& C_{M,\alpha} = C_{M,\alpha}^{\alpha^2} \alpha^2+C_{M,\alpha}^\alpha \alpha+C_{M, \alpha}^0 \\
& C_{M, \delta e}=c_e \delta e \\
& C_T^{\alpha^3}=\beta_1 \phi+\beta_2 \\
& C_T^{\alpha^2}=\beta_3 \phi+\beta_4 \\
& C_T^\alpha=\beta_5 \phi+\beta_6 \\
& C_T^0=\beta_7 \phi+\beta_8
\end{align}
Note that the model is made up of twelve aerodynamic coefficients, which are nonlinear functions of angle of attack and elevator deflection. The thrust of the scramjet $T$ is highly dependent on the states $M_a$ and $\alpha$, as well as the input $\phi$. The relationship is nearly cubic with respect to the angle of attack, and each coefficient in this polynomial is a linear function of $\phi$. The eight coefficients, $\beta_i$, vary with dynamic pressure and altitude. However, since these parameters change at a much slower rate compared to $\phi$ and $\alpha$, the values of $\beta_i$ are assumed to be constant in this study. Readers may refer to \cite{2007parker_control} for more details of this ABHV model. The limits of operational parameters are given in Table~\ref{tab:curvefitting}. Figures~\ref{aerodata}(a-c) show the variation of aerodynamic coefficients $C_L$, $C_D$, and $C_M$ with respect to $\alpha$ at three different elevator settings. Clearly, the slopes of the $C_M - \alpha $ curves show that the aircraft is statically unstable throughout. The variation of thrust with respect to the fuel-equivalence ratio is shown in Fig.~\ref{aerodata}(d). 
\begin{table}
\caption{Operational Range of Parameters\cite{2007parker_control}}
    \centering
    \begin{tabular}{c c c}
    \hline \hline
    \textbf{Parameter}  & \textbf{Lower Bound}  & \textbf{Upper Bound} \\
    
    \hline \hline     $h$ & $25\mathrm{~km}$ & $40\mathrm{~km}$ \\
    \hline     $\phi$& $0.1$ & $1.2$\\
    \hline     $\delta_e$& $-15\mathrm{~deg}$ &$15\mathrm{~deg}$ \\
    \hline     $\alpha$& $-10\mathrm{~deg}$ & $10\mathrm{~deg}$\\
    \hline     $\bar{q}$& $500\mathrm{~lb \cdot ft^{-2}}$ & $2000\mathrm{~lb \cdot ft^{-2}}$\\
         \hline \hline
    \end{tabular}
    
    \label{tab:curvefitting}
\end{table}
\begin{table}
\caption{Aerodynamic Coefficients Values\cite{2007parker_control}}
    \centering
    \begin{tabular}{c c| c c| c c}
    \hline 
    \textbf{Lift}  & \textbf{Value}  & \textbf{Drag} & \textbf{Value} & \textbf{Moment} & \textbf{Value} \\
    
    \hline \hline $C_L^\alpha$ & 
    $4.6773\times10^0\mathrm{~rad^{-1}}$ &
    $C_D^{\alpha^2}$ &
    $5.8224\times 10^0\mathrm{~rad^{-2}}$ &
    $C_{M,\alpha}^{\alpha^2}$ &
    $6.2926\times 10^0\mathrm{~rad^{-2}}$ \\
    \hline

    $C_L^{\delta e}$ & 
    $7.6224\times10^{-1}\mathrm{~rad^{-1}}$ &
    $C_D^{\alpha}$ &
    $-4.5315\times 10^{-2}\mathrm{~rad^{-1}}$ &
    $C_{M,\alpha}^{\alpha}$ &
    $2.1335\times 10^0\mathrm{~rad^{-1}}$ \\
    \hline     

    $C_L^{0}$ & 
    $-1.8714\times10^{-2}\mathrm{}$ &
    $C_D^{\delta e^2}$ &
    $8.1993\times 10^{-1}\mathrm{~rad^{-2}}$ &
    $C_{M,\alpha}^{0}$ &
    $1.8979\times 10^{-1}\mathrm{}$ \\
    \hline     

     & 
     &
    $C_D^{\delta e}$ &
    $2.7699\times 10^{-4}\mathrm{~rad^{-1}}$ &
    $c_e$ &
    $-1.2897\times 10^{0}\mathrm{~rad^{-1}}$ \\
    \hline 

     & 
     &
    $C_D^{0}$ &
    $1.0131\times 10^{-2}\mathrm{}$ &
     &
    \\
    \hline 
            \hline 
    \end{tabular}
    
    \label{tab:curvecoefff}
\end{table}
\begin{table}
\caption{Thrust Coefficients Values\cite{2007parker_control}}
    \centering
    \begin{tabular}{c c c}
    \hline 
    \textbf{Coefficient}  & \textbf{Value}  & \textbf{Units} \\
    
    \hline \hline     $\beta_1$ & $-3.7693\times10^{5}$ & $\mathrm{~lb \cdot ft^{-1} \cdot rad^{-3}}$ \\
    \hline     $\beta_2$&  $-3.7225\times10^{4}$ & $\mathrm{~lb \cdot ft^{-1} \cdot rad^{-3}}$\\
    \hline     $\beta_2$&  $-3.7225\times10^{4}$ & $\mathrm{~lb \cdot   ft^{-1} \cdot rad^{-3}}$\\
    \hline     $\beta_3$&  $2.6814\times10^{4}$ & $\mathrm{~lb \cdot ft^{-1} \cdot rad^{-2}}$ \\
    \hline     $\beta_4$&  $-1.7277\times10^{4}$ & $\mathrm{~lb \cdot ft^{-1} \cdot rad^{-2}}$\\
    \hline     $\beta_5$&  $3.5542\times10^{4}$ & $\mathrm{~lb \cdot ft^{-1} \cdot rad^{-1}}$\\
    \hline     $\beta_6$&  $-2.4216\times10^{3}$ & $\mathrm{~lb \cdot ft^{-1} \cdot rad^{-1}}$\\
    \hline     $\beta_7$&  $6.3785\times10^{3}$ & $\mathrm{~lb \cdot ft^{-1}}$\\
    \hline     $\beta_8$&  $-1.0090\times10^{2}$ & $\mathrm{~lb \cdot ft^{-1}}$\\
     
         \hline \hline
    \end{tabular}
    
    \label{tab:curvethrustcoeff}
\end{table}
Furthermore, to understand the effect of thrust on aircraft stability, the total moment coefficient ($C_{M,total}$) was analyzed at different values of $\phi$ by fixing the dynamic pressure constant at the trim condition given in Table~\ref{tab:Trimcod}. From Eq.(7), the moment coefficient can be written as:
\begin{align}
   & C_{M,total} = z_TT/\frac{1}{2}\rho(M_av^2)S\bar{c} + [C_{M,a} + C_{M,\delta_e}] \\
   & C_{M,total} = C_{M,thrust} + C_{M,aero}
\end{align}
where,
\begin{align}
   & C_{M,thrust} = z_T(C_T^{\alpha^3} \alpha^3+C_T^{\alpha^2} \alpha^2+C_T^\alpha \alpha+C_T^0 )/\frac{1}{2}\rho(M_av^2)S\bar{c}\\
   &   C_{M,thrust} = z_T((\beta_1 \phi+\beta_2)\alpha^3+(\beta_3 \phi+\beta_4) \alpha^2+(\beta_5 \phi+\beta_6) \alpha+ (\beta_7 \phi+\beta_8) )/\frac{1}{2}\rho(M_av^2)S\bar{c}\\
    & C_{M,aero} = C_{M,\alpha}^{\alpha^2} \alpha^2+C_{M,\alpha}^\alpha \alpha+C_{M, \alpha}^0 + c_e\delta_e
\end{align}
As $\phi$ increases, the propulsive moment changes significantly, leading to an increase in $C_{M,total}$ as shown in Fig.~\ref{fig:CMtotalvsaero}. The difference between $C_{M,aero}$ and $C_{M,total}$ becomes more pronounced as $\alpha$ increases. 
\begin{table}
\caption{Nominal Trim Condition}
    \centering
    \begin{tabular}{c c|c c}
    \hline \hline
    \textbf{State}  & \textbf{Value} &\textbf{Input} & \textbf{Value} \\
    \hline \hline
         $h$ & $25.9 \mathrm{~km}$ &$\phi$ & 0.2680 \\
         $M_a$& $6.84 $ & $\delta_e$ & $11.4635 \mathrm{~deg}$\\
         $\alpha$& $1.5153 \mathrm{~deg}$ & \vspace{0 cm}  &\vspace{0 cm}  \\
         $\theta$& $1.5153 \mathrm{~deg}$ & \vspace{0 cm} &\vspace{0 cm} \\
         $Q$& $0 \mathrm{~deg.s^{-1}}$ &\vspace{0 cm} &\vspace{0 cm} \\
         \hline \hline
    \end{tabular}
    
    \label{tab:Trimcod}
\end{table}
To check the veracity of the ABHV model, a modal analysis is carried out first around a cruise flight condition given in Table~\ref{tab:Trimcod}. The eigenvalues/open-loop poles of the linearized model computed independently are given in Table~\ref{tab:eigenvalue}. The corresponding eigenvectors are given in Table~\ref{tab:eigvectors}. The aircraft exhibits two typical modes including an unstable first order short-period like mode, dominated by $\alpha$ and $Q$ (as seen from magnitudes of eigenvectors corresponding to eigenvalues $\lambda_1$ and $\lambda_2$) and a stable phugoid mode dominated by significant variation in $M_a$ and $\theta$ (as seen from magnitudes of eigenvectors corresponding to eigenvalues $\lambda_3$ and $\lambda_4$). The unstable short-period mode can be attributed in part to the positive ${C_M}_\alpha$ (Figs.~\ref{aerodata}(c),~\ref{fig:CMtotalvsaero}). The phugoid mode at this flight condition is stable and very lightly damped. Clearly, the open-loop dynamics of ABHV is unstable. Additionally, the transmission zeros are computed with respect to the input $u = [\phi,\delta_e]$ and the output vector $y = [V,\gamma]$, where $\gamma = \theta -\alpha$ and shown in Fig.~\ref{fig:pole}. Since one of the transmission zeros is on the right-hand side of the complex plane, it exhibits non-minimum phase behavior. These results match with the results in \cite{2007parker_control}, thus confirming the veracity of the model used in this study. Figure~\ref{fig:OL15} shows the vehicle's nonlinear response starting from an initial condition chosen around the operating condition. The simulation results exhibit divergent behavior confirming the operating condition is unstable corroborating conclusions from the modal analysis.
\begin{table}[H]
\caption{Eigenvalues of Linearized System}
    \centering
\begin{tabular}{ccc}
\hline \hline &Eigenvalues & Mode \\
\hline \hline$\lambda_1$ & $1.6944+0.0000 \mathrm{i}$ & Short-period \\
\hline$\lambda_2$ & $-1.7640+0.0000 \mathrm{i}$ & Short-period \\
\hline$\lambda_3$ & $-0.0008 + 0.0063 \mathrm{i}$ & Phugoid \\
\hline$\lambda_4$ & $-0.0008 - 0.0063 \mathrm{i}$ & Phugoid \\
\hline \hline
\end{tabular}
    \label{tab:eigenvalue}
\end{table}
\begin{table}[H]
\caption{Eigenvectors of the Linearized System}
    \centering
    \begin{tabular}{|c| c |c |c|}
    \hline \hline
    \textbf{State}  & \textbf{$\lambda_1$} &\textbf{$\lambda_2$} & \textbf{$\lambda_3,\lambda_4$} \\
    \hline \hline
         $M_a$& $0.0032 + 0.0000i $ & $-0.0026+0.0000i $ & $-0.9761\pm0.0000i$\\
         $\alpha$& $-0.4387 + 0.0000i$ & $-0.4568+0.0000i$  & $-0.0029\mp 0.0000i $  \\
         $\theta$& $-0.4567+0.0000i$ & $-0.4387+0.0000i$ & $0.0243\pm0.2159i $ \\
         $Q$& $-0.7739+0.0000i$ & $0.7739+0.0000i$& $-0.0014\mp0.0000i $\\
         \hline \hline
    \end{tabular}
    
    \label{tab:eigvectors}
\end{table}
In the following, parametric dynamics of the nonlinear model of ABHV are presented. Bifurcation analysis and continuation theory methodology is used for the purpose. A brief overview of the bifurcation analysis based approach is presented next.  
\section{Bifurcation Analysis and Continuation Theory Methodology}
Bifurcation analysis enables a systematic examination of the qualitative changes in the behavior of a nonlinear dynamical system. One of the primary objectives of bifurcation analysis is to create a map that illustrates how the system’s steady states vary with respect to a control parameter, and to provide insights into the nature and stability of these solutions. Loss of stability and originating new type/number of steady state signifying departure behavior, such as, jump, are typical outcomes of a bifurcation analysis. The approach involves solving the full set of simultaneous nonlinear equations of motion of the system, which can be analytically complex. The equations are solved numerically using a numerical continuation tool, which employs a path-following algorithm to trace the system’s steady states from an initial user-provided steady state. The varying control parameter in numerical continuation is known as the continuation parameter, which for an ABHV model can be parameters, for example, elevator deflection, fuel equivalence ratio or center-of gravity position. 

In this paper, bifurcation analysis and continuation methods were implemented using the continuation and bifurcation software AUTO-07p \cite{2009Deodel}. A brief introduction to the continuation algorithm follows.
Consider a general autonomous dynamical system of the form:
\begin{equation}
    \dot{x} = f(x,u_c),  ~~ x,f\in R^n
    \label{eqn:xdot}
\end{equation}
where '$f$'is a vector of '$n$' smooth (differentiable) functions, $x$ is the state vector and $u_c$ is the input vector. 
 The system is in equilibrium when: 
\begin{equation}
    \dot{x} = 0
    \label{xzero}
\end{equation}
A continuation algorithm computes steady states of Eq.~\ref{eqn:xdot} by solving a simultaneous set of nonlinear algebraic equations as a function of a varying parameter known as continuation parameter of the system $u\in u_c$, while the remaining parameters of the system $p\in u_c$ remain fixed in a continuation.
\begin{equation}
    \dot{x} = f(x,u,p) = 0
    \label{eqn:px}
\end{equation}
Further, the continuation algorithm evaluates the stability of each of these trim states. This is done by numerically computing the Jacobian matrix,
\begin{equation}
    J  = \frac{df}{dx}|_{x_0}
    \label{eqn:px1}
\end{equation}
and its eigenvalues, where ‘$x_0$’ denotes the equilibrium state (solution of Eq. (22)) at which the Jacobian matrix is evaluated.
The maps generated by solving the Eqs.~\ref{eqn:px} along with ~\ref{eqn:px1} are known as bifurcation diagrams. 
Numerical continuation necessitates the availability of at least one solution, which users can acquire through various methods. One common approach is the time-integration method, where the system described in Eq.~\ref{eqn:xdot} is simulated for a sufficiently long duration until the states converge to their final values, assuming a stable solution exists nearby. Alternatively, Newton's method can be employed to determine equilibria. It is worth noting that in numerous published works, the terms 'bifurcation analysis' and 'numerical continuation' are often used interchangeably. 

'Bifurcation' refers to change in stability/number/type of steady states which occurs at the critical points where one or more eigenvalues migrate across the imaginary axis in the complex plane. Conditions are evaluated at these critical points to identify the type of bifurcation. The type of bifurcation of steady states encountered in aircraft flight dynamics fall into one of the following three categories\cite{2005paranjape_airplane}:
\begin{enumerate}
   \item  Limit Point (LP)/Fold Bifurcation: This phenomenon occurs when a single real eigenvalue crosses the imaginary axis at the origin. It can be understood as the convergence of two solution branches, one stable and the other unstable, that meet and cancel each other out. Alternatively, it can be conceptualized as a solution branch reversing its direction, hence earning the moniker of a turning point or, more precisely, a saddle-node bifurcation point. This bifurcation type plays a crucial role in generating dynamic behaviors such as jumps and hysteresis.
 
   \item  Branch Point (BP): Similar to the Limit Point Bifurcation, a Branch Point occurs when a single real eigenvalue crosses the imaginary axis at the origin. Depending on the nonlinearities, two different types of branching happen at a BP: transcritical and pitchfork. In a transcritical bifurcation, two solution branches intersect and exchange stability. For cubic and higher-odd nonlinearities in the system, super or subcritical pitchfork bifurcation is possible, in which a single solution branch bifurcates into multiple odd solution branches or vice versa. A subcritical pitchfork bifurcation may also lead to jump and hysteresis for a higher order system.

   \item Hopf Bifurcation (HB): This occurs when a pair of complex conjugate eigenvalues migrates across the imaginary axis, leading to the emergence of a branch of limit cycles. Hopf bifurcations can be categorized as either supercritical or subcritical, depending on the order of nonlinearities. At this bifurcation, origination of periodic steady-states: $x(t) = x(t+T), T>0$ becomes possible. These are special types of periodic states in the sense that trajectories converge to or diverge from these, also known as Limit cycles, depending on their stability properties. Stability of limit cycle is determined by computing Floquet multipliers. Bifurcations of limit cycles similar to bifurcations of equilibrium states are also possible depending on how they lose stability, characterized by movement of Floquet multipliers across the unit circle in complex plane; for a more detailed description readers may refer to. Advanced continuation algorithms such as AUTO-07p also computes branches of Limit cycles. 
  
\end{enumerate}

\section{Results and Discussions}
This section presents the results obtained from the bifurcation analysis of the nonlinear ABHV model. The results are presented in the form of bifurcation diagrams in which steady-state (trim) values of a state variable are plotted against a varying (control) parameter. The open-loop dynamics is explored to understand the behavior of the system. Insights into the unstable nature of the vehicle are derived using bifurcation analysis and time simulations. 

\textbf{Case 1: Elevator continuation with fuel-equivalence ratio fixed at }\textbf{$\phi=0.2680$}

\begin{figure}[H]
\centering
    \subfloat[]{
        \includegraphics[width=0.47\textwidth]{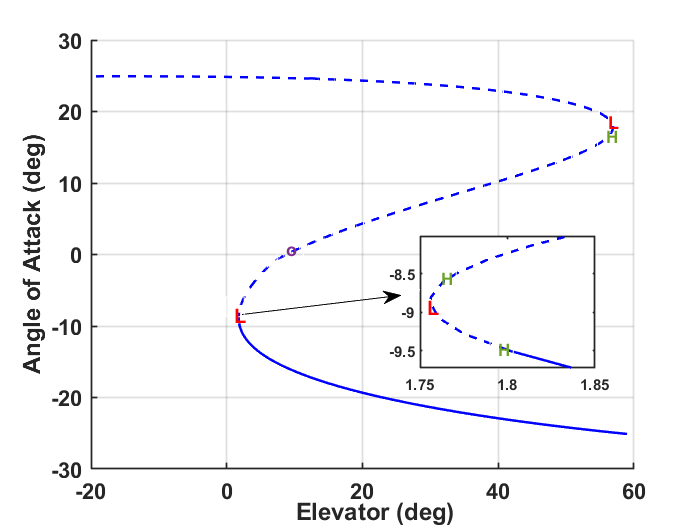}
        \label{fig:C1}
    }
    \hfill
    \subfloat[]{
        \includegraphics[width=0.47\textwidth]{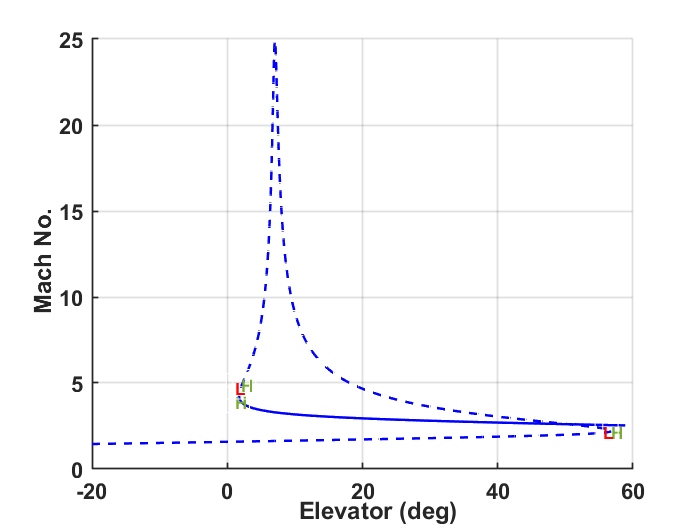}
        \label{fig:C2}
    }
        \hfill
    \subfloat[]{
        \includegraphics[width=0.47\textwidth]{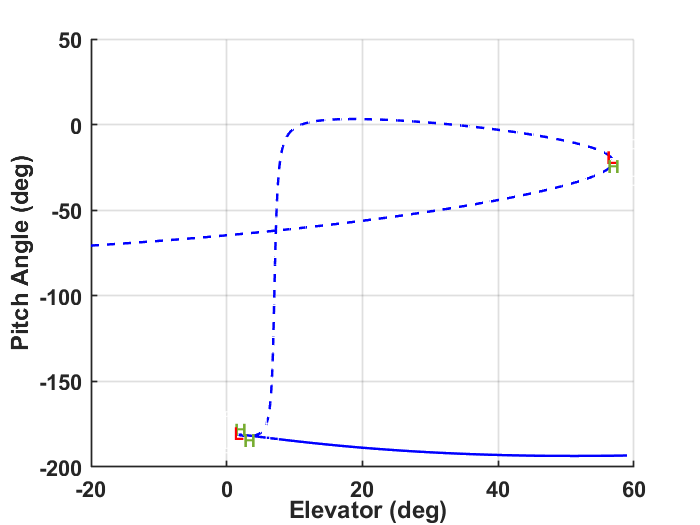}
        \label{fig:C3}
    }
        \hfill
    \subfloat[]{
        \includegraphics[width=0.47\textwidth]{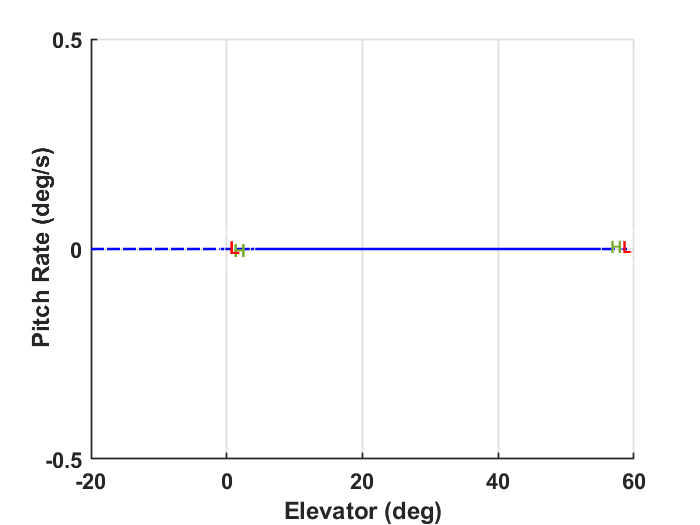}
        \label{fig:C4}
    }
    \caption{Open-Loop Bifurcation Diagram $-$ Elevator Continuation (dashed line- unstable solutions, solid line - stable solutions, 'H' = Hopf point, 'L' = fold point).}
    \label{Open}
\end{figure}

Figure~\ref{Open} shows equilibrium solution branches of the ABHV model as function of the elevator deflection for a fixed value of fuel-equivalence ratio. Solid lines represent stable equilibrium solutions and dashed lines represent branch of unstable equilibrium solutions. Initial observation indicates that all solutions in the specified angle-of-attack range (-10 to +10 deg) in Table~\ref{tab:curvefitting} are unstable as shown in Fig.~\ref{fig:C1}. The point 'o' is the starting point of the continuation which is the nominal trim condition given in Table~\ref{tab:Trimcod}. Three branches of equilibrium solutions can be seen in the elevator range (1.75 to 15 deg) with solution branches folding at elevator deflection of 1.75 deg. Elevator deflection below the fold point results in a jump at the fold point, leading to jump in angle-of-attack. The range of elevator deflection stretched beyond limits specified in Table~\ref{tab:curvefitting} is primarily to compute all connecting solution branches in a continuation from the given starting point.

The upper equilibrium solution branch available for elevator deflection below 1.75 deg post-jump is unstable and hence not practically accessible. In particular, there are no possible solutions for the negative elevator deflection. Also, the high angle-of-attack stable solution branch below the fold point value of elevator deflection is beyond specified range of angle-of-attack for which the model is valid. ABHV's conventional configuration is inherently unstable as reported in literature (\cite{2005bolender_non}\cite{2007bolender_nonlinear}) which is also true for this model. Solid lines representing stable equilibrium solutions with large values of pitch angle refer to conditions that may become possible in a loss-of-control scenario. Presence of Hopf bifurcation around the fold point indicates possible existence of oscillatory behavior which can be verified by carrying out numerical simulation. The unstable feasible solutions need to be stabilized using perhaps an appropriate stability augmentation system. Branch of solutions with lower angle-of-attack, acceptable pitch angle values and Mach number appropriate for the engine, however unstable, are only important from an operation point of view.
Figure~\ref{fig:C2} depicts the variation of Mach number with respect to elevator deflection. The turning point is observed at Mach 4.26. The stable branch emanating from this point nearly saturates around Mach 3 for most variations in elevator deflection. However, the unstable branch extends up to Mach 25 before descending to approximately Mach 5 as the elevator varies. Interestingly, this unstable branch also corresponds to a positive lower angle of attack, which is a usually acceptable operating flight condition. 

In Fig.~\ref{fig:C3}, the pitch angle solutions indicate a stable branch with very high values, highlighting an unrealistic flight scenario. On the other hand, an unstable solution branch corresponds to acceptable flight conditions. The pitch rate bifurcation diagram shown in Fig.~\ref{fig:C4} indicates that all equilibrium solutions have zero pitch rate, which is naturally one of the equilibrium constraints in longitudinal flight condition. 
\begin{figure} [H]
\centering
     \subfloat[]{
        \includegraphics[width=0.47\textwidth, height=0.38\textwidth]{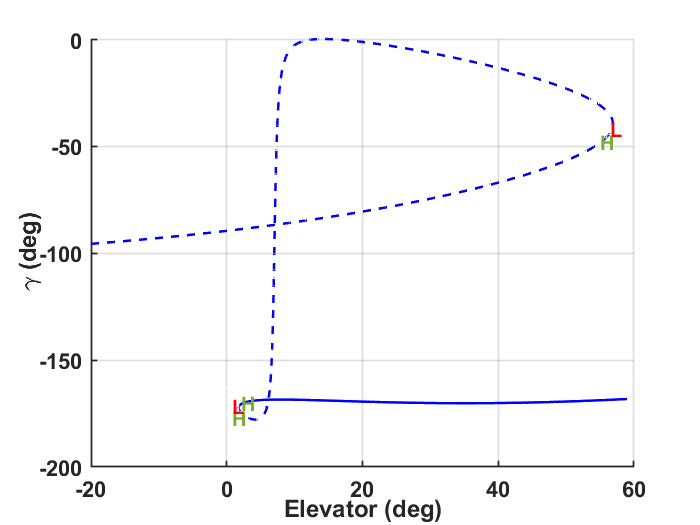}
        \label{fig:gamma_dele}
    }
    \hfill
     \subfloat[]{
        \includegraphics[width=0.50\textwidth]{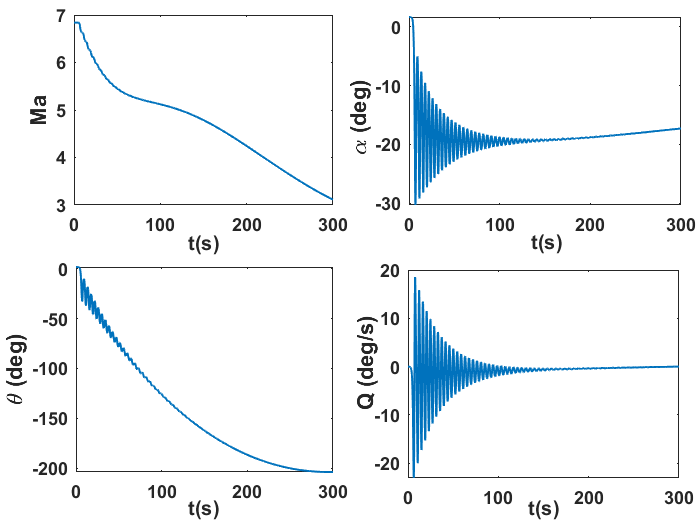}
        \label{fig:OL300_1}
    }
    \caption{(a) Flight path angle variation as elevator varies, (b)Time simulation showing flight variables settling to the stable point at the nominal condition.}
    \label{gamma_dele_open}    
\end{figure}
Study of the above diagrams suggests that all flyable operating conditions are unstable for the ABHV model considered in the paper. Flight path angle, an important parameter from an operational point of view, is also plotted in Fig.~\ref{fig:gamma_dele}. Solutions indicate that for this value of fuel-equivalence ratio, only level flight and descent equilibrium conditions, even though unstable, are possible over a small range of elevator deflection.  

One among many advantages of using bifurcation analysis techniques is that it provides a good guess of initial conditions that can be used for numerical simulation of dynamics of a nonlinear dynamical system model such as ABHV. One such simulation carried out around a stable equilibrium solution corroborating results from bifurcation analysis is presented in Fig.~\ref{fig:OL300_1}. 
 
\textbf{Case 2: Fuel-Equivalence Ratio ($\phi$) Continuation with elevator setting fixed at $\delta_e=12.5~\mathrm{deg}$}

\begin{figure}[H]
\centering
    \subfloat[]{
        \includegraphics[width=0.47\textwidth, height=0.36\textwidth]{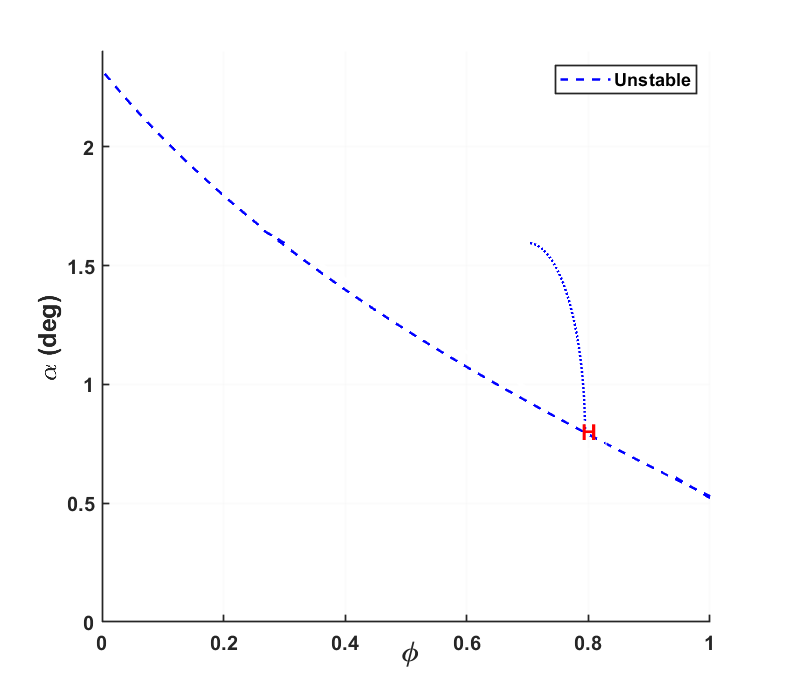}
        \label{fig:D1}
    }
    \hfill
    \subfloat[]{
        \includegraphics[width=0.47\textwidth]{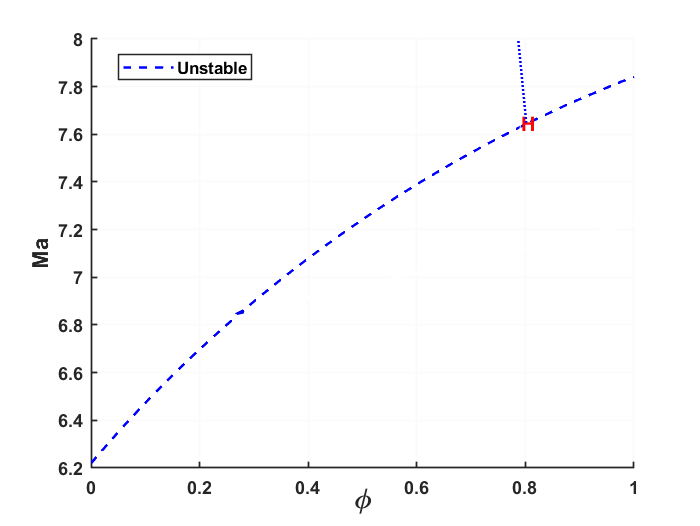}
        \label{fig:D2}
    }
        \hfill
    \subfloat[]{
        \includegraphics[width=0.47\textwidth]{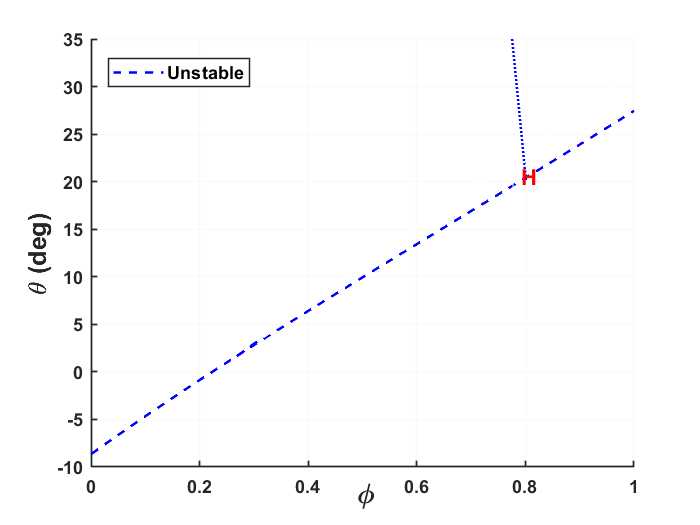}
        \label{fig:D3}
    }
        \hfill
    \subfloat[]{
        \includegraphics[width=0.47\textwidth]{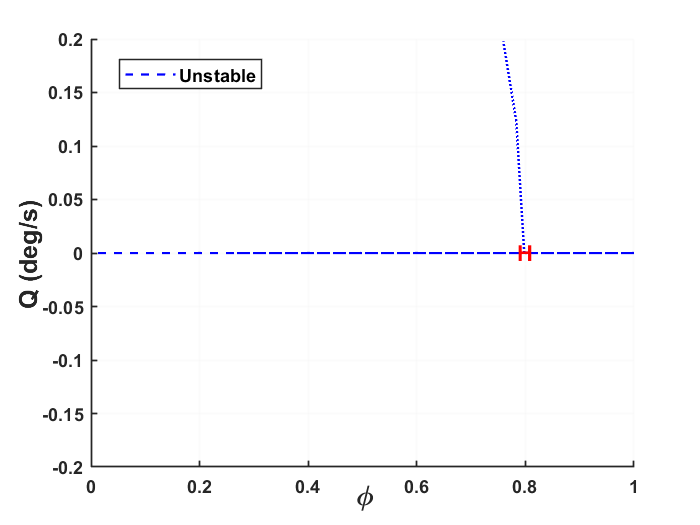}
        \label{fig:D4}
    }
    \caption{Open-Loop Bifurcation Diagram $-$ Fuel-Equivalence Ratio ($\phi$) continuation.}
    \label{fig:phi}
\end{figure}

Figure~\ref{fig:phi} illustrates the locus of equilibrium solutions as the fuel-equivalence ratio ($\phi$) varies. The cruise flight condition at Mach 6.84 given in Table~\ref{tab:curvefitting} is chosen as the starting equilibrium solution for continuation to run. The elevator value is maintained at a constant value of $12.5\mathrm{deg}$. This starting point is marked as ‘o’ in Fig.~\ref{Open} and corresponds to $\phi = 0.268$, which is used in the elevator continuation in Case 1. The dashed lines in Fig.~\ref{fig:phi} are unstable solutions and the dotted lines are periodic solutions emerging from the Hopf bifurcation point marked ‘H’ in Fig.~\ref{fig:phi}. Initial observations from these diagrams indicate that the behavior of ABHV is unstable throughout the range of fuel-equivalence ratio for this particular elevator setting.

\begin{figure}
\centering
     \subfloat[]{
        \includegraphics[width=0.50\textwidth]{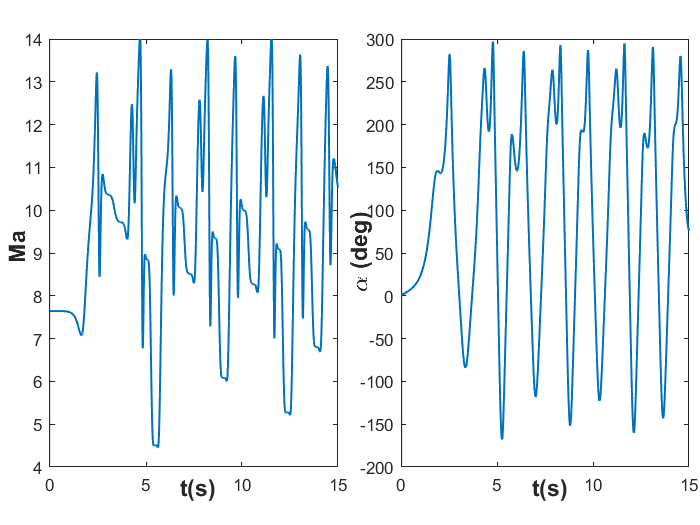}
        \label{fig:3a}
    }
    \hfill
    \subfloat[]{
        \includegraphics[width=0.47\textwidth, height=0.38\textwidth]{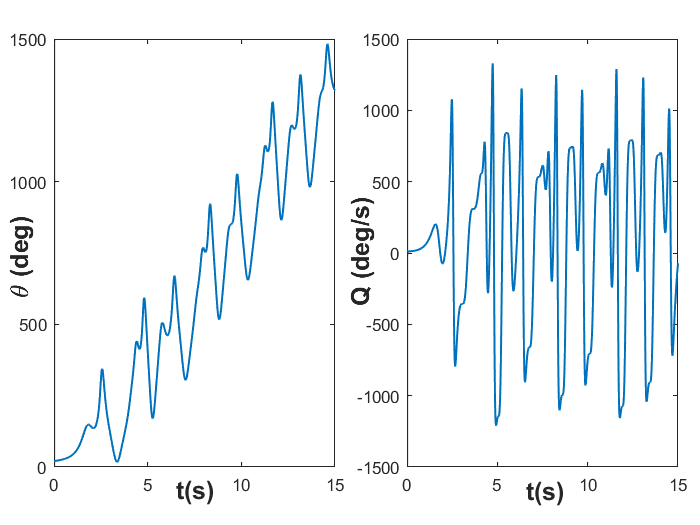}
        \label{fig:3b}
    }
    \caption{Time Simulation near the Hopf point at $\phi=0.8$, illustrating an aperiodic response.}
    \label{fig:3}    
\end{figure}

\begin{figure}[H]
\centering
     {
\includegraphics[width=0.55\textwidth]{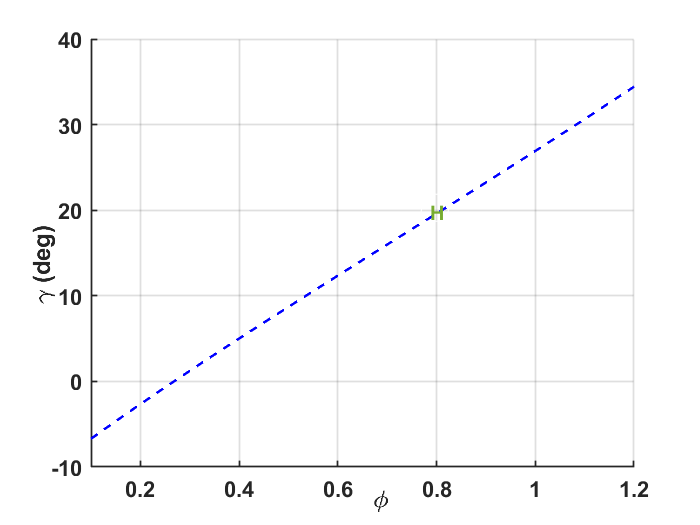}        
        \caption{ Flight-path angle ($\gamma $) in  $\phi$ continuation.}
        \label{fig:gamma_phi}
    }
\end{figure}

At $\phi=0.8$, a Hopf bifurcation is detected due to possible brief migration of a pair of eigenvalues to the left half complex plane and immediate return. Continuation to track limit cycles and their bifurcations beyond the Hopf bifurcation point is an additional feature of AUTO-07p\cite{2009Deodel}. Results for the continuation for periodic orbits are appended to the branch of equilibrium solution from the first continuation for steady solutions, both results are presented together in Fig.~\ref{fig:phi}. The extended branch of unstable periodic solutions emanating from the Hopf point 'H' in Fig.~\ref{fig:phi} suggests a subcritical type of Hopf bifurcation that often gives rise to existence of multiple period solutions in the vicinity and further to higher order oscillations leading to possibly chaotic oscillations. Close examinations of dynamics in the vicinity of the Hopf bifurcation point require computation of transients. 
The time simulation results presented in Fig.~\ref{fig:3} around the Hopf point exhibit erratic oscillations in the angle of attack, Mach number, and pitch rate, while the pitch angle diverges continuously. This verifies the typical behavior of subcritical Hopf bifurcation. Flight path angle, an important parameter from operational point of view is plotted in Fig.~\ref{fig:gamma_phi}. At this elevator deflection, changing fuel-equivalence ratio appears to cover a wide range of flight path angle including descent, level, and climb flight conditions. However, the whole range of unstable trims clearly indicates the need for integrated flight-propulsion feedback.  

\textbf{Case 3: A continuation near zero angle-of-attack solution with elevator fixed at $\delta_e=8.79~\mathrm{deg}$}

\begin{figure}
\centering
    \subfloat[]{
        \includegraphics[width=0.47\textwidth]{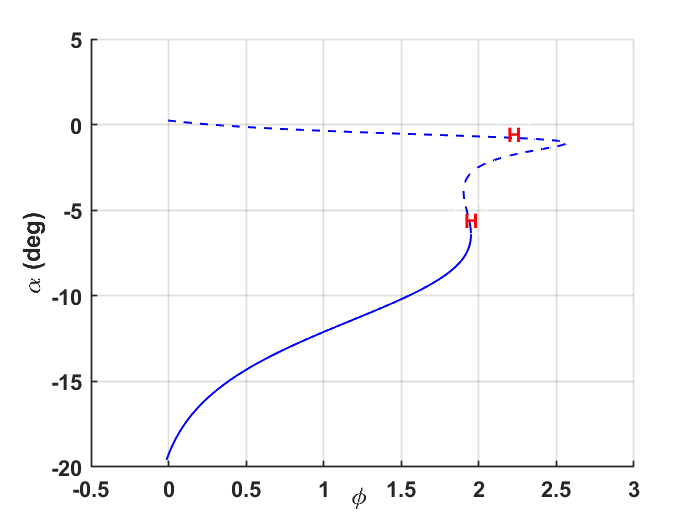}
        \label{fig:D11}
    }
    \hfill
    \subfloat[]{
        \includegraphics[width=0.47\textwidth]{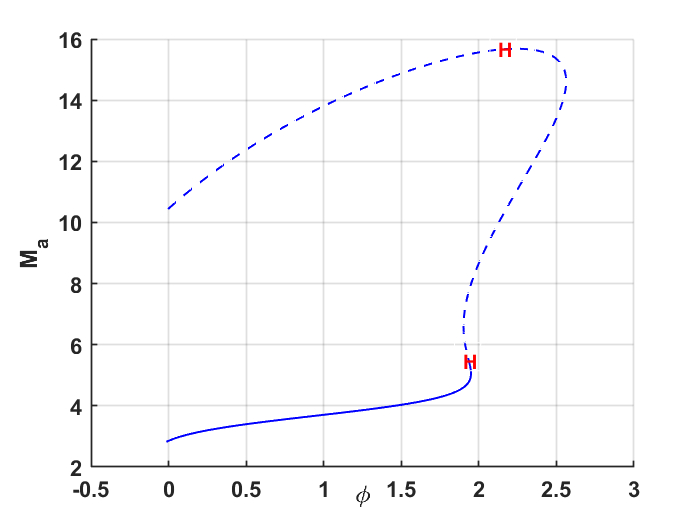}
        \label{fig:D21}
    }
        \hfill
    \subfloat[]{
        \includegraphics[width=0.47\textwidth]{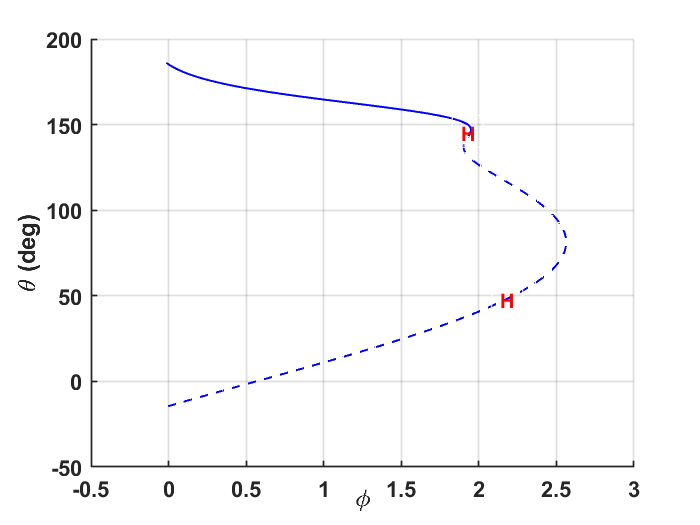}
        \label{fig:D31}
    }
        \hfill
    \subfloat[]{
        \includegraphics[width=0.47\textwidth]{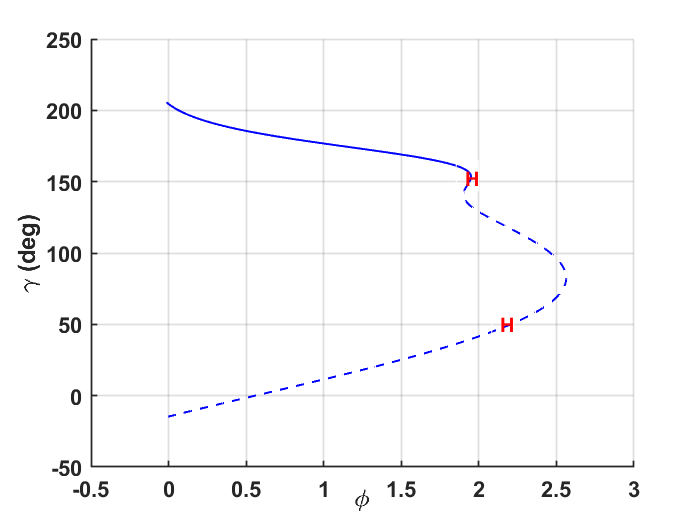}
        \label{fig:D41}
    }
    \caption{Open-Loop Bifurcation Diagram $-$ Fuel-Equivalence Ratio ($\phi$) continuation around $\alpha = 0$ deg.}
    \label{fig:alpha0}
\end{figure}

A further continuation is carried out to examine equilibrium solutions at small angles of attack with changing fuel-equivalence ratio. Starting point for this continuation is selected from the first continuation result plotted in Fig.~\ref{Open} and elevator is fixed at the respective value. Bifurcation diagrams for this case are presented in Figs.~\ref{fig:alpha0}. As noted from the diagrams, a pair of equilibrium solutions exist for most values of fuel-equivalence ratio for this elevator deflection, with unstable low angle-of-attack solutions and moderately high angle-of-attack unstable solutions. Flight path angle ($\gamma$) goes from a negative value (descent flight) at $\phi = 0$, passing through level flight at nearly $\phi = 0.5$, and eventually going into positive value (climb flight) beyond $\phi = 0.5$. Looking at the Mach number plot, one can notice that as fuel-equivalence ratio is increased from the Mach number increase for this elevator deflection, however, bringing back
the fuel-equivalence ratio to zero does not bring the Mach number to zero. Presence of Hopf bifurcation points can be also noted, but beyond the acceptable limits of $\phi = 1.2$. Much of the operationally acceptable solution branch (small-$\alpha$, small-$\gamma$) correspond to very high Mach number and are unstable.   

\textbf{Case 4: Level flight trim computation and analysis}

It is to be noted that all above diagrams (in Cases 1-3) correspond to either a fixed value of the fuel-equivalence ratio and varying elevator deflection (Case 1) or a fixed value of elevator with varying fuel-equivalence ratio (Cases 2 and 3). These parametric variations lead to different kinds of longitudinal trim, e.g. different flight path angles which are operationally not always meaningful. Instead, a more realistic scenario where the deployment of more than one control parameter is mandatory needs to be investigated. Investigation of, for example, level flight (represented by zero flight path angle, zero pitch rate, constant non-zero Mach number) trims, where states of aircraft are constrained and which requires scheduling of both fuel-equivalence ratio and elevator, may highlight more useful information from the performance and stability point of view. Many examples of such actual flights defined by constraints and their bifurcation analysis using a special technique known as Extended Bifurcation Analysis (EBA) have been reported in the literature\cite{2001ananthkrishnan_level,2005paranjape_airplane}. In the first step of EBA, setting the constraint equation for level flight trim ($\gamma =0$) and appending it with the equations for longitudinal trims (Eqs.(1-4)) result in the augmented set of equations:
\begin{align}
   & f(x,u)= 0 \\
    & \gamma=0
    \label{augmented}
\end{align}
\begin{figure}[H]
\centering
     {
        \includegraphics[width=0.55\textwidth,height=0.45\textwidth]{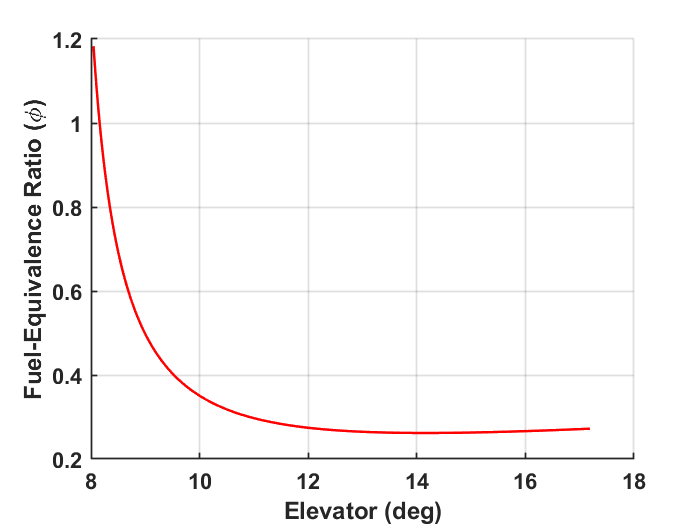}
        \caption{ $\delta_e - \phi $ variation - EBA continuation for level-flight.}
         \label{fig:deltae-phi}
    }
\end{figure}  
Now, fuel-equivalence ratio can be varied along with the elevator to satisfy the constraint. This variation obtained by running a continuation of the augmented set of equations (Eq.(\ref{augmented})) is shown in Fig.~\ref{fig:deltae-phi}. The variation of fuel-equivalence ratio with elevator deflection is a functional schedule that would be required to fly different level flight trims. Interestingly, it reveals a pareto front, suggesting combination of higher fuel-equivalence ratio at lower elevator deflections and lower fuel-equivalence ratio at higher elevator deflection to fly level flight trims. This can be easily corroborated analytically using the equation of longitudinal trim given by $C_m =0$, which is a function of both fuel-equivalence ratio as well as elevator deflection. Increase in fuel-equivalence ratio amounts to a proportional increase in thrust and thereby the increase in moment produced as a result. The elevator deflection must be decreased to counteract the increased moment, and vice versa. An optimal combination can be read directly as the point where distance between the origin of the plot and the point on the curve is minimum. 
\begin{figure}[H]
\centering
     \subfloat[]{
        \includegraphics[width=0.50\textwidth,height = 0.34\textwidth]{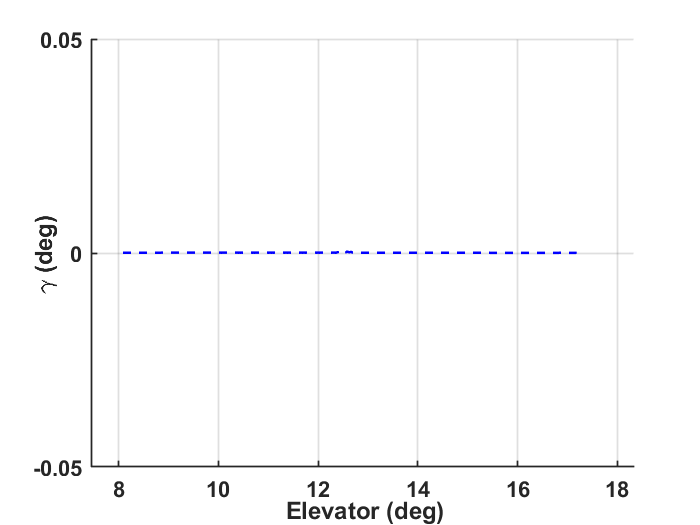}
        \label{fig:OL300}
        
    }\hfill
    \subfloat[]{
        \includegraphics[width=0.47\textwidth,height = 0.34\textwidth]{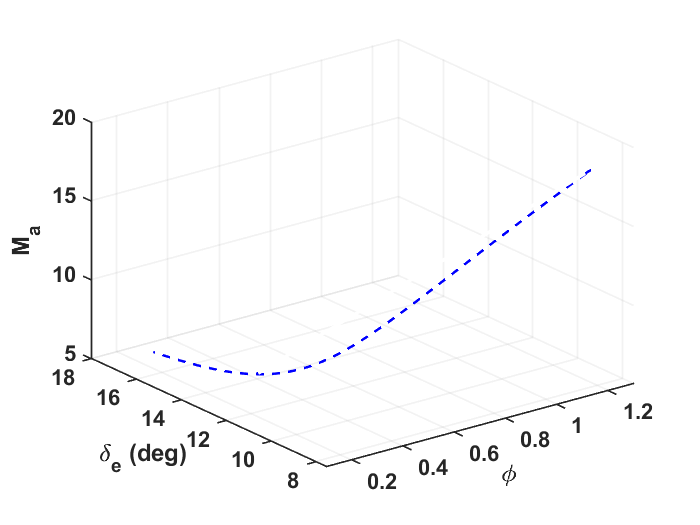}
        \label{fig:alpha-phi}
     
    }
    \caption{ (a)Flight path angle, (b) Mach number variation as $\delta_e - \phi $ vary for level-flight constraint. }
    \label{EBA2}    
\end{figure}
\begin{figure}[H]
\centering
    \subfloat[]{
        \includegraphics[width=0.47\textwidth]{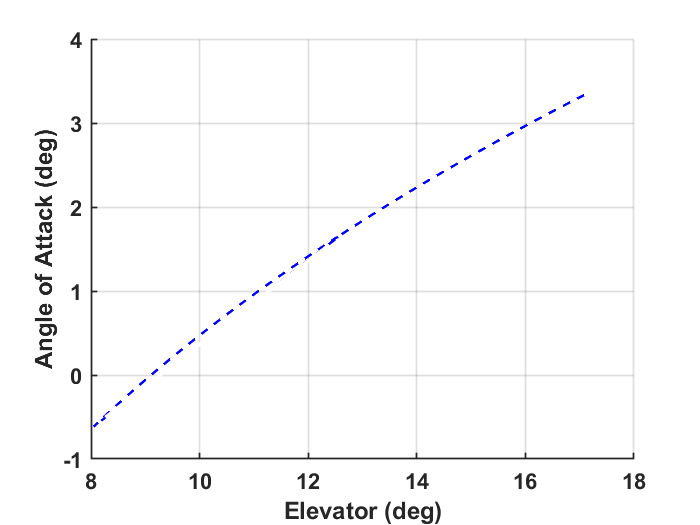}
        \label{fig:3D}
       
    }
    \hfill
    \subfloat[]{
        \includegraphics[width=0.47\textwidth]{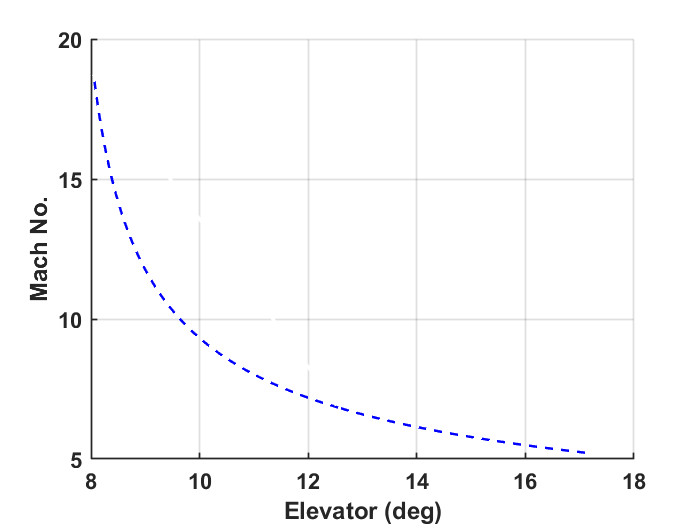}
        \label{fig:phidele}
      
    }
      \hfill
    \subfloat[]{
        \includegraphics[width=0.47\textwidth]{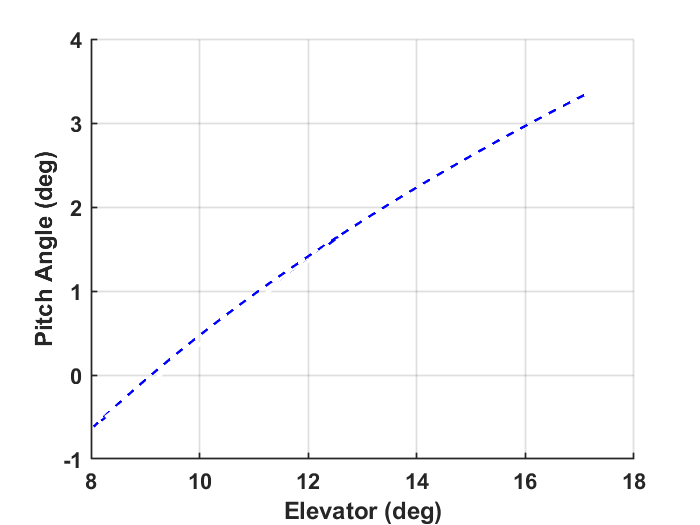}
        \label{fig:C33}
    }
        \hfill
    \subfloat[]{
        \includegraphics[width=0.47\textwidth]{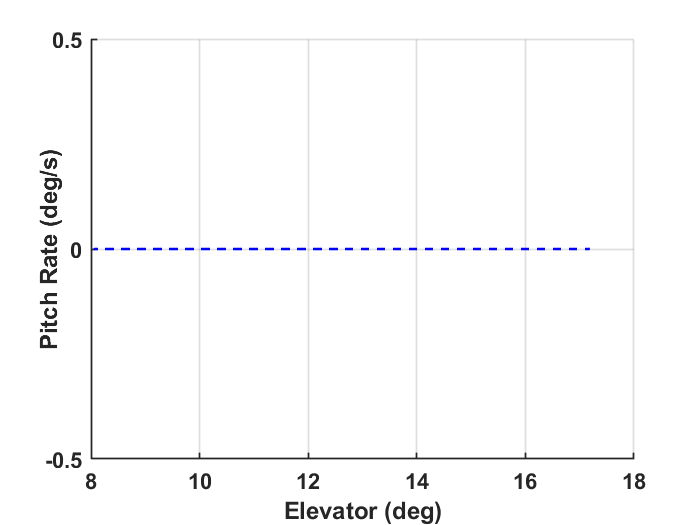}
        \label{fig:C44}
    }
    \caption{Locus of level-flight trims obtained from EBA continuation. }
    \label{two}
\end{figure}
 In order to capture stability of these (level) flight trims, a second continuation with fuel-equivalence ratio varying with elevator as in Fig.~\ref{EBA2} is carried out, but without the constraint equation now, following the second step of the EBA\cite{2001ananthkrishnan_level}:
 \begin{equation}
         \dot{x} = f(x,u,p_1(u))
 \end{equation}
where, $p_1$ is the parameter that varies as a function of $u$.
 Results of the second continuation are presented in Fig.~\ref{two}. Not surprisingly, all level flight trims are unstable as the vehicle is unstable by design. Several combinations of speed and angle of attack level flight trim can be observed from the bifurcation diagrams for the combination of fuel-equivalence ratio and elevator deflection. Interestingly, a pareto front exists for Mach number – elevator deflection combination also as a direct result of the variation in thrust due to fuel-equivalence ratio. Larger elevator deflection has an adverse effect on Mach number which decreases drastically with slight aerodynamic advantage via increase in angle of attack.

\section{Concluding Remarks}
A model of air-breathing hypersonic vehicle longitudinal dynamics is presented in this paper and is verified using modal analysis carried out around a design operating condition reported in literature. The model is inherently nonlinear due to aerodynamics and coupling between moment produced due to thrust and aerodynamics, and is further investigated using bifurcation analysis and continuation methods for its parametric dynamic behavior. Four cases are presented to identify suitable operating conditions in various flights, including decent, level and climb. Stability and bifurcations are captured and plausible explanations about dynamic behavior of ABHV from bifurcation diagrams are presented. As the model being inherently unstable, most equilibrium solutions are found to be unstable. Oscillatory behavior resulting out of the presence of a Hopf bifurcation point and jump as result of fold bifurcation are identified. Time simulation results are presented, where necessary, to corroborate inferences made out from bifurcation analysis results. Design of a suitable stability augmentation system to stabilize the equilibrium solution branches of interest and further investigation of close-loop dynamic behavior of ABHV is recommended as the future scope of work.  

\bibliography{sample}

@inproceedings{1980mehra_bifurcation,
  title={Bifurcation analysis of aircraft high angle-of-attack flight dynamics},
  author={Mehra, R and Carroll, J},
  booktitle={6th Atmospheric Flight Mechanics Conference},
  address={AIAA Paper 1980-1599},
  year={1980},
  DOI = "10.2514/6.1980-1599"
}

@inproceedings{1992schmidt_dynamics,
  title={Dynamics and control of hypersonic aeropropulsive/aeroelastic vehicles},
  author={Schmidt, David},
  booktitle={Guidance, Navigation and Control Conference},
  address = {AIAA Paper 1992-4326},
  year={1992},
  DOI = "10.2514/6.1992-4326"
}

@inproceedings{1992chavez_integrated,
  title={An integrated analytical aeropropulsive/aeroelastic model for the dynamic analysis of hypersonic vehicles},
  author={Chavez, Frank and Schmidt, David},
  booktitle={Astrodynamics Conference},
  address ={AIAA Paper 1992-4567},
  year={1992},
  DOI = "10.2514/6.1992-4567"
}

@incollection{1993schmidt_problems,
  title={Problems in Control System Design for Hypersonic Vehicles},
  author={Schmidt, DK},
  booktitle={Automatic Control in Aerospace 1992},
  address = {Pergamon},
  pages={89--96},
  year={1993},
  publisher={Elsevier}, 
  DOI = "10.1016/B978-0-08-041715-8.50016-X"
}

@article{1994chavez_analytical,
  title={Analytical aeropropulsive-aeroelastic hypersonic-vehicle model with dynamic analysis},
  author={Chavez, Frank R and Schmidt, David K},
  journal={Journal of Guidance, Control, and Dynamics},
  volume={17},
  number={6},
  pages={1308--1319},
  year={1994},
  DOI = "10.2514/3.21349"
}

@article{1995bilimoria_integrated,
  title={Integrated development of the equations of motion for elastic hypersonic flight vehicles},
  author={Bilimoria, Karl D and Schmidt, David K},
  journal={Journal of Guidance, Control, and Dynamics},
  volume={18},
  number={1},
  pages={73--81},
  year={1995},
  DOI = "10.2514/3.56659"
}

@article{1997avanzini_bifurcation,
  title={Bifurcation analysis of a highly augmented aircraft model},
  author={Avanzini, Giulio and Matteis, Guido de},
  journal={Journal of Guidance, Control, and Dynamics},
  volume={20},
  number={4},
  pages={754--759},
  year={1997},
  DOI = "10.2514/2.4108"
}

@article{1998littleboy_using,
  title={Using bifurcation methods to aid nonlinear dynamic inversion control law design},
  author={Littleboy, DM and Smith, PR},
  journal={Journal of guidance, control, and dynamics},
  volume={21},
  number={4},
  pages={632--638},
  year={1998},
  DOI = "10.2514/2.4282"
}

@article{2000ortega_bifurcation,
  title={Bifurcation analysis of a feedback system with dead zone and saturation},
  author={Ortega, MG and Aracil, J and Gordillo, F and Rubio, FR},
  journal={IEEE Control Systems Magazine},
  volume={20},
  number={4},
  pages={91--101},
  year={2000},
  publisher={IEEE},
  DOI = "10.1109/37.856182"
}

@article{2001ananthkrishnan_level,
  title={Level flight trim and stability analysis using extended bifurcation and continuation procedure},
  author={Ananthkrishnan, Narayan and Sinha, Nandan K},
  journal={Journal of Guidance, Control, and Dynamics},
  volume={24},
  number={6},
  pages={1225--1228},
  year={2001},
  DOI = "10.2514/2.4839"
}

@inproceedings{2005bolender_non,
  title={A non-linear model for the longitudinal dynamics of a hypersonic air-breathing vehicle},
  author={Bolender, Michael and Doman, David},
  booktitle={AIAA guidance, navigation, and control conference and exhibit},
  address={AIAA Paper 2005-6255},
  year={2005},
  DOI = "10.2514/6.2005-6255"
}

@inproceedings{2005paranjape_airplane,
  title={Airplane level turn performance, including stability constraints, using EBAC method},
  author={Paranjape, Aditya and Ananthkrishan, N},
  booktitle={AIAA Atmospheric Flight Mechanics Conference and Exhibit},
  Address ={AIAA Paper 2005-5800},
  year={2005},
  DOI = "10.2514/6.2005-5800"
}

@article{2006richardson_design,
  title={Design of a gain-scheduled flight control system using bifurcation analysis},
  author={Richardson, Thomas and Lowenberg, Mark and DiBernardo, Mario and Charles, Guy},
  journal={Journal of Guidance, Control, and Dynamics},
  volume={29},
  number={2},
  pages={444--453},
  year={2006},
  DOI = "10.2514/1.13902"
}

@article{2007parker_control,
  title={Control-oriented modeling of an air-breathing hypersonic vehicle},
  author={Parker, Jason T and Serrani, Andrea and Yurkovich, Stephen and Bolender, Michael A and Doman, David B},
  journal={Journal of Guidance, Control, and Dynamics},
  volume={30},
  number={3},
  pages={856--869},
  year={2007},
  DOI = "10.2514/1.27830"
}

@article{2007bolender_nonlinear,
  title={Nonlinear longitudinal dynamical model of an air-breathing hypersonic vehicle},
  author={Bolender, Michael A and Doman, David B},
  journal={Journal of spacecraft and rockets},
  volume={44},
  number={2},
  pages={374--387},
  year={2007},
  DOI = "10.2514/1.23370"
}

@inproceedings{2008torrez_effects,
  title={Effects of improved propulsion modelling on the flight dynamics of hypersonic vehicles},
  author={Torrez, Sean and Driscoll, James and Bolender, Michael and Oppenheimer, Michael and Doman, David},
  booktitle={AIAA Atmospheric Flight Mechanics Conference and Exhibit},
  address ={AIAA Paper 2008-6386},
  year={2008},
  DOI = "10.2514/6.2008-6386"
}

@misc{2009Deodel,
author="E. J. Doedel",
title="AUTO-07P: Continuation and bifurcation software for ordinary differential equations</em>,",
journal="Concordia University",
year="2009",
URL="https://cir.nii.ac.jp/crid/1370569869258146569"
}

@article{2011huifeng_control,
  title={Control-oriented modeling for air-breathing hypersonic vehicle using parameterized configuration approach},
  author={Huifeng, LI and Ping, LIN and Dajun, XU},
  journal={Chinese Journal of Aeronautics},
  volume={24},
  number={1},
  pages={81--89},
  year={2011},
  publisher={Elsevier},
  DOI = "10.1016/S1000-9361(11)60010-1"
}

@article{2011falkiewicz_reduced,
  title={Reduced-order aerothermoelastic framework for hypersonic vehicle control simulation},
  author={Falkiewicz, Nathan J and Cesnik, Carlos ES and Crowell, Andrew R and McNamara, Jack J},
  journal={AIAA journal},
  volume={49},
  number={8},
  pages={1625--1646},
  year={2011},
  DOI = "10.2514/1.J050802"
}

@inproceedings{2012crespo_analysis,
  title={Analysis of control strategies for aircraft flight upset recovery},
  author={Crespo, Luis and Kenny, Sean and Cox, David and Murri, Daniel},
  booktitle={AIAA Guidance, Navigation, and Control Conference},
  address ={AIAA Paper 2012-5026},
  year={2012},
  DOI = "10.2514/6.2012-5026"
}

@article{2014zhang_modeling,
  title={Modeling and analysis of an air-breathing flexible hypersonic vehicle},
  author={Zhang, Xi-bin and Zong, Qun},
  journal={Mathematical Problems in Engineering},
  volume={2014, No. 1},
  year={2014},
  publisher={Hindawi},
  DOI = "10.1155/2014/264247"
}

@article{2015xu_overview,
  title={An overview on flight dynamics and control approaches for hypersonic vehicles.},
  author={Xu, Bin and Shi, ZhongKe},
  journal={Sci. China Inf. Sci.},
  volume={58},
  number={7},
  pages={1--19},
  year={2015},
  DOI = "10.1007/s11432-014-5273-7"
}

@article{2015sharmanumerical,
  title={Numerical continuation and bifurcation analysis in aircraft design: an industrial perspective},
  author={Sharma, Sanjiv and Coetzee, Etienne B and Lowenberg, Mark H and Neild, Simon A and Krauskopf, Bernd},
  journal={Philosophical Transactions of the Royal Society A: Mathematical, Physical and Engineering Sciences},
  volume={373},
  number={2051},
  pages={20140406},
  year={2015},
  publisher={The Royal Society Publishing}, 
  DOI = "10.1098/rsta.2014.0406"
}

@article{2018sudalagunta_aeroelastic,
  title={Aeroelastic control-oriented modeling of an airbreathing hypersonic vehicle},
  author={Sudalagunta, Praneeth Reddy and Sultan, Cornel and Kapania, Rakesh K and Watson, Layne T and Raj, Pradeep},
  journal={Journal of Guidance, Control, and Dynamics},
  volume={41},
  number={5},
  pages={1136--1149},
  year={2018},
  publisher={American Institute of Aeronautics and Astronautics},
  DOI = "10.2514/1.G002777"
}

@article{2020wu_analysis,
  title={Analysis and recovery of aircraft deep-stall phenomena using bifurcation analysis},
  author={Wu, Dawei and Chen, Mou and Ye, Hui},
  journal={IEEE Access},
  volume={8},
  pages={29319--29333},
  year={2020},
  publisher={IEEE},
  DOI = "10.1109/ACCESS.2020.2972570"
}

@book{sinha2021elementary,
  title={Elementary flight dynamics with an introduction to bifurcation and continuation methods},
  author={Sinha, Nandan K and Ananthkrishnan, N},
  year={2021. Chap 8}, 
  publisher={CRC Press}
}

@article{2022singh_longitudinal,
  title={Longitudinal Trim and Stability Analysis of Generic Air-Breathing Hypersonic Vehicle using Bifurcation Method},
  author={Singh, Ritesh and Prakash, Om and Joshi, Sudhir and Jeppu, Yogananda},
  journal={INCAS Bulletin},
  volume={14},
  number={3},
  pages={111--123},
  year={2022},
  publisher={INCAS-National Institute for Aerospace Research" Elie Carafoli"},
  DOI = "10.13111/2066-8201.2022.14.3.10 "
}

@article{1982carroll_bifurcation,
  title={Bifurcation analysis of nonlinear aircraft dynamics},
  author={Carroll, James V and Mehra, Raman K},
  journal={Journal of Guidance, Control, and Dynamics},
  volume={5},
  number={5},
  pages={529--536},
  year={1982},
  DOI = "10.2514/3.56198"
}

@article{2001charles_bifurcation,
  title={Bifurcation tailoring of equilibria: a feedback control approach},
  author={Charles, G and DI BERNARDO, Mario and Lowenberg, M and Stoten, DP and Wang, XF and others},
  journal={Latin American Applied Research},
  volume={31},
  number={3},
  pages={199--210},
  year={2001}
}

\end{document}